\documentstyle[12pt]{article}
         \makeatletter
         \def\thefigure{\@arabic\c@figure}\def\fps@figure{tbp}
         \def\ftype@figure{1}\def\ext@figure{lof}
         \def\fnum@figure{\protect\footnotesize Fig.\ \thefigure}
         \def\thetable{\@arabic\c@table}
         \def\fps@table{tbp}\def\ftype@table{2}\def\ext@table{lot}
         \def\fnum@table{\protect\footnotesize Table \thetable}
         \def\@listI{\leftmargin\leftmargini\parsep=0pt\itemsep=0pt}
         \def\thebibliography#1{\section{References}\vspace*{-10pt}\list
          {[\arabic{enumi}]}{\settowidth\labelwidth{[#1]}\leftmargin\labelwidth
          \advance\leftmargin\labelsep
          \usecounter{enumi}}
          \def\newblock{\hskip .11em plus .33em minus .07em}
          \sloppy\clubpenalty4000\widowpenalty4000
          \sfcode`\.=1000\relax}
         
         \def\@nomath#1{\ifmmode \fi}
         \def\mmycite{\@ifnextchar [{\@tempswatrue\@mmycitex}
             {\@tempswafalse\@mmycitex[]}}
         \def\@mmycitex[#1]#2{\if@filesw\immediate%
         \write\@auxout{\string\citation{#2}}\fi
           \def\@citea{}\@mmycite{\@for\@citeb:=#2\do
             {\@citea\def\@citea{,}\@ifundefined
                {b@\@citeb}{{\bf ?}\@warning
                {Citation `\@citeb' on page \thepage \space undefined}}%
         \hbox{\csname b@\@citeb\endcsname}}}{#1}}
         \def\@mmycite#1#2{{{\scriptsize#1}\if@tempswa , #2\fi}}
         \def\mycite#1{$^{\protect\mmycite{#1}}$}
         \def\@cite#1#2{{#1\if@tempswa , #2\fi}}
         \def\thesection {\arabic{section}}
         \def\section#1{\addtocounter{section}{1}\setcounter{subsection}{0}
              \bigskip\medskip{\noindent\bf\thesection.\ #1}
              \medskip}
         \def\thesubsection {\arabic{section}.\arabic{subsection}}
         \def\subsection#1{\addtocounter{subsection}{1}
              \medskip{\noindent\thesubsection.\ #1}
              \medskip}
         \makeatother
\begin{document}
\begin{flushright} LBL-35688 \end{flushright}
\vspace*{0.3in}
\begin{center}
  {\bf PARTON EQUILIBRATION IN ULTRARELATIVISTIC HEAVY ION
COLLISIONS\footnote{Collaborations with my colleagues are gratefully
acknowledged. This work was supported by the Director, Office of Energy
Research, Division of Nuclear Physics of the Office of High
Energy and Nuclear Physics of the U.S. Department of Energy
under Contract No. DE-AC03-76SF00098.}}\\
  \bigskip
  \bigskip
  Xin-Nian Wang\\
  {\em Nuclear Science Division, MS 70A-3307\\
       Lawrence Berkeley Laboratory, Berkeley, CA 94720, USA}\\
  \bigskip
\end{center}
\smallskip
{\footnotesize
\centerline{ABSTRACT}
\begin{quotation}
\vspace{-0.10in}
Medium effects which include color screening and Landau-Pomeranchuk-Migdal
suppression of induced radiation are discussed in connection with the
equilibration of dense partonic system in ultrarelativistic heavy ion
collisions. Taking into account of these medium effects, the equilibration
rate for a gluonic gas are derived and the consequences are discussed.
\end{quotation}}

\section{Introduction}

Strong interactions involved in hadronic collisions can be generally
divided into two categories depending on the scale of momentum transfer
$Q^2$ of the processes. When $Q^2\sim\Lambda^2_{QCD}$, the collisions
are nonperturbative in QCD and are considered soft.  These interactions
can be approximated by some effective theories in which partons inside
a nucleon cannot be resolved and nucleons interact coherently by the exchange
of mesons or soft Pomerons. These kind of coherent interactions will result
in the collective excitations as observed in experiments at low and
intermediate energies, $\sqrt{s}<$ a few GeV. On the other hand, if $Q^2$ is
much larger than $\Lambda^2_{QCD}$, parton model becomes relevant and
they interact approximately incoherently. Due to the high $Q^2$,
parton interactions can be calculated via perturbative QCD (pQCD), given
the initial parton distribution functions inside a nucleus. At high
collider energies as RHIC and LHC, the hard processes of parton
interactions become dominant and heavy ion collisions are generally very
violent\mycite{BLAIZ}. In these violent collisions, it is quite likely that
colors
are liberated during the collisions and we can start with a color deconfined
system.

   In the last few years, many efforts have been made to estimate the
initial parton density by perturbative
calculations\mycite{eskola,wang1,geiger,wang2}.
Though the results depend on parton structure functions inside a nucleus
and the modeling of the underlying soft interactions, a general consensus
is that a dense partonic system will be produced during the early stage
of ultrarelativistic heavy ion collisions. However, the system is not
necessarily in both thermal and chemical equilibrium. The initial
system is totally dominated by gluons and the quark and anti-quark
densities are far below their chemical equilibrium values.
Numerical simulations
of parton cascade can be made with certain simplifications. However, the
complexity of these calculations makes it difficult to obtain a clear
understanding of the different time scales on various parameters and
model assumptions. Most importantly, medium effects and interferences
are difficult to incorporate. In this talk, I would like to describe
two medium effects. Color screening is caused by the interaction of
a color charge with the medium which gives a natural infrared cut-off
to regularize the parton cross sections inside the dense system\mycite{screen}.
Another medium effect is the so-called Landau-Pomeranchuk-Migdal effect of
the reduced radiation by multiple scattering through the medium\mycite{XWMG}.
These medium effects can be utilized to obtain a parameter-free
set of equations, based on perturbative QCD in a dense partonic
medium, that describes the evolution of quark and gluon distributions
towards equilibrium\mycite{therm}.

\section{Color Screening}

        Let us first look at the color screening in the initially produced
parton system. Following the standard calculation(in Coulomb gauge)
of screening in the time-like gluon propagator in a medium
of gluonic excitation, we have the screening mass\mycite{screen},

\begin{equation}
        \mu^2_D=-\frac{3\alpha_s}{\pi^2}\lim_{|{\bf q}|\rightarrow 0}
        \int d^3k \frac{|{\bf k}|}{{\bf q}\cdot{\bf k}}
        {\bf q}\cdot{\bf\nabla_k}f({\bf k}).
\end{equation}
Now, instead of using the Bose-Einstein distribution for
the thermalized case, we simply relate the phase space
density $f({\bf k})$ to the initial gluon distribution
calculated from pQCD. At high energies we find that the transverse
and longitudinal screening lengths are very close. We assume
that the gluon momentum distribution is similar to a thermal
distribution
\begin{equation}
        f_g(k)=\lambda_ge^{-u\cdot k/T},
\end{equation}
with $\lambda_g$ and $T$ characterizing the thermalization of
the gluon gas. With this distribution, one obtain the effective
color screening mass
\begin{equation}
       \mu_D^2=\lambda_g g^2T^2. \label{eq:scr}
\end{equation}
We will use this effective color screening mass as an infrared
cut-off for parton interaction cross sections inside a non-equilibrium
gluonic gas.

\section{Induced Radiation}

The leading processes which contribute to the gluon equilibration
are those with induced gluon radiation.
Let us consider the simplest case of induced radiation from
a two-quark scattering. The Born amplitude for two-quark
scattering $(p_i,k_i)\rightarrow (p_f,k_f)$ through one gluon
exchange is,
\begin{equation}
{\cal M}_{el}=ig^2T^a_{AA'}T^a_{BB'}\frac{\bar u(p_f)\gamma_{\mu}u(p_i)
        \bar u(k_f)\gamma^{\mu}u(k_i)}{(k_i-k_f)^2},
        \label{eq:el1}
\end{equation}
where $A$, $A'$, $B$, and $B'$ are the initial and final
color indices of the beam and target partons. The corresponding
elastic cross section is,
\begin{equation}
\frac{d\sigma_{el}}{dt}=C^{(1)}_{el}\frac{\pi\alpha_s^2}{s^2}
                         2\frac{s^2+u^2}{t^2},
\end{equation}
where $C^{(1)}_{el}=C_F/2N=2/9$ is the color factor for a single
elastic quark-quark scattering and $s$,$u$, and $t$ are the Mandelstam
variables.

Taking into account the dominant contribution, the radiation
amplitude from a single scattering is,
\begin{eqnarray}
{\cal M}_{rad}&\equiv& i\frac{{\cal M}_{el}}{T^a_{AA'}T^a_{BB'}}
                    {\cal R}_1, \nonumber \\
{\cal R}_1&\simeq&2ig\vec{\epsilon}_{\perp}\cdot\left[
\frac{{\bf k}_{\perp}}{k^2_{\perp}}+\frac{{\bf q}_{\perp}-{\bf k}_{\perp}}
{({\bf q}_{\perp}-{\bf k}_{\perp})^2}\right]T^a_{AA'}[T^a,T^b]_{BB'},
\label{eq:rad2}
\end{eqnarray}
where ${\cal M}_{el}$ is the elastic amplitude as given in Eq.~(\ref{eq:el1}),
and ${\cal R}_1$ is defined as the radiation amplitude induced by a single
scattering. For later convenience, all the color matrices are included in
the definition of the radiation amplitude ${\cal R}_1$.  With the above
approximations, we then recover the differential
cross section for induced gluon bremsstrahlung by a single collision as
originally derived by Gunion and Bertsch~\mycite{GUNION},
\begin{equation}
\frac{d\sigma}{dtdyd^2k_{\perp}}=\frac{d\sigma_{el}}{dt}
\frac{dn^{(1)}}{dyd^2k_{\perp}},
\end{equation}
where the spectrum for the radiated gluon is,
\begin{equation}
\frac{dn^{(1)}}{dyd^2k_{\perp}}\equiv\frac{1}{2(2\pi)^3C^{(1)}_{el}}
\overline{\left|{\cal R}_1\right|^2}
=\frac{C_A\alpha_s}{\pi^2}\frac{q^2_{\perp}}
{k^2_{\perp}({\bf q}_{\perp}-{\bf k}_{\perp})^2}. \label{eq:spec1}
\end{equation}
In the square modulus of the radiation amplitude, an average and a sum
over initial and final color indices and polarization are understood.
The above formula is also approximately valid for induced radiation
off a gluon line, except that the color factor $C^{(1)}_{el}$ in the
elastic cross section has to be replaced by $1/2$ for $gq$ and $9/8$
for $gg$ scatterings.

One nonabelian feature in the induced gluon radiation amplitude,
Eq.~(\ref{eq:rad2}), is the singularity at ${\bf k}_{\perp}={\bf q}_{\perp}$
due to induced radiation along the direction of the exchanged gluon.
For $k_{\perp}\ll q_{\perp}$, we note that the induced radiation
from a three gluon vertex can be neglected as compared to the leading
contribution $1/k^2_{\perp}$. However, at large $k_{\perp}\gg q_{\perp}$,
this three gluon amplitude is important to change the gluon spectrum
to a $1/k^4_{\perp}$ behavior, leading to a finite average transverse
momentum. Therefore, $q_{\perp}$ may serve as a cut-off for $k_{\perp}$
when one neglects the amplitude with the three gluon vertices as we
will do when we consider induced radiation by multiple scatterings
in the next section.

\section{EFFECTIVE FORMATION TIME}

The radiation amplitude induced by multiple scatterings has been
discussed in Ref.~[\cite{XWMG}]. We here only briefly discuss the
case associated with double scatterings. We consider two static
potentials separated by a distance $L$ which is much larger than
the interaction length, $1/\mu$. For convenience of discussion we
neglect the color indices in the case of an abelian interaction first.
The radiation amplitude associated with double scatterings is,
\begin{equation}
{\cal R}_2^{\rm QED}=ie\left[\left(\frac{\epsilon\cdot p_i}{k\cdot p_i}
-\frac{\epsilon\cdot p}{k\cdot p}\right)e^{ik\cdot x_1}+\left(
\frac{\epsilon\cdot p}{k\cdot p}-\frac{\epsilon\cdot p_f}{k\cdot p_f}
\right)e^{ik\cdot x_2}\right],\label{eq:radQED}
\end{equation}
where $p=(p_f^0,p_z,{\bf p}_{\perp})$ is the four-momentum of
the intermediate parton line which is put on mass shell by the pole
in one of the parton propagators, $x_1=(0,{\bf x}_1)$, and
$x_2=(t_2,{\bf x}_2)$ are the four-coordinates of the two potentials
with $t_2=(z_2-z_1)/v=Lp^0/p_z$. We notice that the amplitude has
two distinguished contributions from each scattering. Especially,
the diagram with a gluon radiated from the intermediate
line between the two scatterings contributes both as the final
state radiation for the first scattering and the initial state
radiation for the second scattering.  The relative phase factor
$k\cdot(x_2-x_1)=\omega(1/v-\cos \theta)L$ then will determine the
interference between radiations from the two scatterings.
If we define the formation time as
\begin{equation}
\tau(k)=\frac{1}{\omega(1/v-\cos \theta)}\simeq
\frac{2\omega}{k^2_{\perp}},
\end{equation}
then Bethe-Heitler limit is reached when $L\gg \tau(k)$. In this
limit, the intensity of induced radiation is simply additive in the
number of scatterings. However, when $L\ll\tau(k)$, the final
state radiation from the first scattering completely cancels
the initial state radiation from the second scattering. The
radiation pattern looks as if the parton has only suffered a single
scattering. This is often referred to as the Landau-Pomeranchuk-Migdal
(LPM) effect\mycite{LPM}. The corresponding limit is usually called
factorization
limit.

The extrapolation to the general case of
$m$ number of scatterings gives us the radiation amplitude in QCD,
\begin{equation}
{\cal R}_m=i2g\frac{\vec{\epsilon}\cdot{\bf k}_{\perp}}{k^2_{\perp}}
T^{a_1}_{A_1A_1'}\cdots T^{a_m}_{A_mA_m'}\sum_{i=1}^m
\left(T^{a_m}\cdots[T^{a_i},T^b]\cdots T^{a_1}\right)_{BB'}e^{ik\cdot x_i},
\label{eq:radm}
\end{equation}
which contains $m$ terms each having a common momentum dependence in the high
energy limit, but with different color and phase factors. The above
expression should also be valid for a gluon beam jet, with the corresponding
color matrices replaced by those of an adjoint representation.
The spectrum of soft bremsstrahlung associated with multiple
scatterings in a color neutral ensemble is, similar to Eq.~(\ref{eq:spec1}),
\begin{equation}
\frac{dn^{(m)}}{dyd^2k_{\perp}}=\frac{1}{2(2\pi)^2C^{(m)}_{el}}
\overline{|{\cal R}_m|^2}\equiv C_m(k)\frac{dn^{(1)}}{dyd^2k_{\perp}},
\label{eq:specm}
\end{equation}
where $C^{(m)}_{el}=(C_F/2N)^m$ is the color factor for
the elastic scattering cross section without radiation. $C_m(k)$,
defined as the ``radiation formation factor'' to characterize the
interference pattern due to multiple scatterings, can be expressed as,
\begin{equation}
C_m(k)=\frac{1}{C^m_FC_AN}\sum_{i=1}^m\left[C_{ii}
+2Re\sum_{j=1}^{i-1}C_{ij}e^{ik\cdot(x_i-x_j)}\right],
\end{equation}
where the color coefficients are defined as
\begin{equation}
C_{ij}=Tr\left(T^{a_m}\cdots [T^b,T^{a_i}]\cdots T^{a_1}T^{a_1}\cdots
[T^{a_j},T^b]\cdots T^{a_m}\right).
\end{equation}

If we average over the interaction points
${\bf x}_i$ according to a linear kinetic theory, we find that
an effective formation time in QCD can be defined as,
\begin{equation}
\tau_{\rm QCD}(k)=r_2\tau(k)=\frac{C_A}{2C_2}\frac{2\cosh y}{k_{\perp}},
                            \label{eq:ftime}
\end{equation}
which depends on the color representation of the jet parton. The
induced gluon radiation due to multiple scattering will be suppressed
when the mean free path $\lambda$ is much smaller than the effective
formation time.

\section{Equilibration Rates}

The most important reactions for establishing gluon equilibrium are
$gg\leftrightarrow ggg$. Elastic
scattering processes, on the other hand, are crucial for maintaining local
thermal equilibrium. Multi-gluon radiation is presumably suppressed by
color screening, while radiative processes
involving quarks have smaller cross sections due to QCD color factors.

The evolution of the gluon density $n$ according to
reactions mentioned above can be described by a rate equation.
Adding the equation for energy conservation assuming only longitudinal
expansion we end up with a closed set of equations determining the
temperature $T(\tau )$ and the gluon ``fugacity''
$\lambda _g(\tau )\equiv n/n_{eq}(T)$
as a function of the proper time $\tau $,
\begin{eqnarray}
  {\dot\lambda_g\over\lambda_g} + 3{\dot T\over T} + {1\over\tau} &=
&R_3 (1-\lambda_g),\\
\lambda_g^{3/4} T^3\tau &=& \hbox{const.}
\end{eqnarray}
This set of evolution equations is completely controlled by the gluon
production rate, $R_3=\frac{1}{2}\sigma_3n$.

Taking into account of LPM effect, the cross section
for $gg\rightarrow ggg$ processes can be written as,
\begin{equation}
\frac{d\sigma_3}{d^2q_{\perp}dyd^2k_{\perp}}=\frac{d\sigma_{el}}{d^2q_{\perp}}
\frac{dn^{(1)}}{d^2k_{\perp}dy}
k_{\perp}\cosh y\theta(\lambda-\tau_{\rm QCD}(k))
\theta(E-k_{\perp}\cosh y), \label{eq:ds3}
\end{equation}
where $\tau_{\rm QCD}(k)$ is given by Eq.~(\ref{eq:ftime}), the second
$\theta$-function is for energy conservation. The gluon
density distribution induced by a single scattering is given by
Eq.~(\ref{eq:spec1}) which must be regularized by the color screening
mass $\mu_D$ in Eq.~(\ref{eq:scr}). Using the elastic cross section of
gluon scattering regularized also by $\mu_D$, we obtain a fugacity
independent mean free path
\begin{equation}
\lambda^{-1} = \sigma^{2\to 2} n = {\textstyle{9\over 8}}
a_1 \alpha_sT, \label{eq:mfp}
\end{equation}

Using these values we evaluate the chemical gluon equilibration rate
$R_3 = {1\over 2} n\sigma_3$, as defined in Eq. (\ref{eq:ds3}), numerically.
This rate scales with the temperature linearly but is a complicated
function of the gluon fugacity.  The result
can approximated by an analytical fit,
\begin{equation}
R_3 = 2.1 \alpha_s^2 T \left( 2\lambda_g - \lambda_g^2\right)^{1/2},
\label{50}
\end{equation}
which will be used in solving the time dependent rate equations.
The rate we thus obtained has a nonlinear dependence on the fugacity due
to the inclusion of the LPM effect and the effective color screening mass.
If LMP effect were not included, the rate whould be much larger, thus
would lead to a fast equilibration.

\end{document}